\documentclass[12pt]{article}
\usepackage{geometry}                		
\usepackage{graphicx}				
\usepackage{amsmath}								
\usepackage{amssymb}
\usepackage{amscd}
\usepackage{yfonts}
\usepackage{amsmath,amsthm,amssymb}
\usepackage{amsthm}

\newcommand {{\bx}} {\bf x}
\newcommand {{\bk}} {\bf k}

\date{}

\medskip

\title{Asymptotic commutativity in Jordan  algebras}
\author{Albert Schwarz}
\date{}							

\begin{document}
\author {  A. Schwarz\\ Department of Mathematics\\ 
University of 
California \\ Davis, CA 95616, USA,\\ schwarz @math.ucdavis.edu}


\maketitle
\begin{abstract}
We estimate commutators of quadratic operators $Q_a$ in Jordan algebras. These estimates can be used to construct the scattering theory in quantum fields theories formulated in terms of Jordan algebras.
\end{abstract}
{\bf Keywords} Jordan algebra; quadratic operators; scattering theory
\vskip .1in

The standard algebraic approach to quantum  theory is based on associative algebras with involution. To develop scattering theory in this approach one should fix the action of translation group on the algebra and impose some conditions of asymptotic commutativity (in the case of bosons) or asymptotic anticommutativity (in the case of fermions). In recently suggested geometric approach to quantum theory (\cite {SG}-\cite {GA4AR}) Jordan algebras are more natural than associative algebras. It was shown in \cite {GA4AR} that one can develop scattering theory in the framework of Jordan algebras  if quadratic operators $Q_a=2R_a^2-R_{a^2}$ asymptotically commute.  ( Here $R_a$ 
 stands for the multiplication operator: $R_a(b)=b\circ a.$ )
 
We prove  asymptotic commutativity of operators $Q_a$ under certain conditions. This result was used in \cite{GA4AR} to prove the existence of (inclusive) scattering matrix. (The present paper will be published as an appendix to \cite{GA4AR}.)

Let us start with some remarks in the simple case when Jordan algebra $\cal B$  is obtained as a set of self-adjoint elements of associative algebra $\cal A$ with involution $^*$ and the operation $a\circ b=\frac 1 2 (ab+ba).$ We assume that $\cal A$ is a C*-algebra, then one says that $\cal B$ is a  JC-algebra. It is easy to check that in the case at hand  we have $Q_a=l(a)r(a)$ where $l(a)$ and $r(a)$ are the operators of multiplication by $a$ from the left and from the right in $\cal A.$
Using  the relations $[l(a),l(b)]=
l([a,b]), [r(a),r(b)]=r([b,a])$ and $Q_aQ_b=l(ab)r(ba), Q_bQ_a=l(ba)r(ab)$ , we obtain the estimate 
\begin {equation}
\label {c}
||[Q_a,Q_b]||\leq 2||a||\cdot||b||\cdot ||[a,b]||.
\end {equation}
Similarly one can estimate the commutator $[Q_a,Q_b]$ in terms of anticommutator of $a,b$ (or equivalently in terms of Jordan product $a\circ b$):
\begin {equation}
\label {a}
||[Q_a,Q_b]||\leq 2||a||\cdot||b||\cdot ||[a,b]_+||=4 ||a||\cdot||b||\cdot ||a\circ b||.
\end {equation}

The proof remains almost the same: we should use the relations $l(ab)= -l(ba)+l([a,b]_+), r(ba)=-r(ab)+r([a,b]_+).$

Let us suppose that commutative Lie group  (translation group) acts on $\cal A$ by means of involution-preserving automorphisms $\alpha ({\bx}, \tau)$; we use the notation $\alpha ({\bx},\tau)a=a({\bx},\tau)$. Then the  same group acts on $\cal B$; it transforms  the operators  $Q_a$  into the operators  $Q_{a({\bx},\tau)}.$

We say that the elements $a,b\in \cal A$ asymptotically commute if 
$$[a({\bx},\tau),b]\leq\rho({\bx},\tau)$$ where $\rho$ is small for large $\bx$ (for definiteness we assume that that $\rho$ tends to zero faster than any power of $||\bx||$ and has at most polynomial growth with respect to $\tau$)

The elements $a,b\in\cal A$ asymptotically anticommute if a similar relation is valid for anticommutator
$$[a{(\bx},\tau),b]_+\leq\rho({\bx},\tau).$$

It follows from (\ref {c}), (\ref {a}) that in both cases operators $Q_a, Q_b$ asymptotically commute:

$$||[Q_a({\bx},\tau),Q_b]||\leq 2||a||\cdot ||b||\rho({\bx},\tau).$$

We would like to generalize the above relations to the case of Jordan Banach algebras (JB-algebras). The possibility of such a generalization is prompted by the following statements:

a) If $a, b$ are two elements of JB-algebra $\cal B$ and $a\circ b=0$ then operators $Q_a$ and $Q_b$ commute  \cite {ACP}, \cite {SHE}.

b) If $a,b$ are two elements of JB-algebra $\cal B$ and operators $R_a,R_b$ commute than the operators $Q_a$ and $Q_b$ also commute \cite {WET}.

 One says that $a,b\in \cal B$  operator commute iff the operators $R_a$ and $R_b$ commute.
For $JB$-algebras it was proven in \cite {WET} that  the operator commutativity of $a,b$ implies that  $a^2$ operator commutes with $b$ and $b^2$ (similarly $b^2$ operator commutes with $a$). This means that the operators $Q_a$ and $Q_b$  commute.

Let us suppose that commutative Lie group  (translation group) acts on $\cal B$ by means of automorphisms $\alpha (\bx, \tau)$. The same group acts on the operators $R_a$ and $Q_a$ transforming them into the operators $R_{\alpha (\bx,\tau)\it a}$ and $Q_{\alpha (\bx,\tau)\it a}.$ We say that $a,b\in \cal B$ asymptotically operator commute if $||[R_{\alpha (\bx,\tau)\it a},R_b]||<.$ It is natural to conjecture that  in this case the operators $Q_a, Q_b$ also asymptotically commute, i. e. $||[Q_{{\alpha ({\bf x},\tau)}{\it a}},Q_b]||<\rho(\bx,\tau).$  The methods of 
\cite {WET} are not sufficient to prove this conjecture. However, {\it  if the pairs $(a,b), (a^2,b), (a,b^2), (a^2,b^2)$ asymptotically
operator commute it is obvious that $Q_a, Q_b$ asymptotically commute}. This statement is sufficient for applications we have in mind.

The same statements are true in  more general case when the translation group acts by structural transformations: $\alpha ({\bx},\tau)\in Strg (\cal B)$, where $Strg (\cal B)$ denotes the structure group (the group generated by automorphisms and invertible operators $Q_a$).

 Let us generalize the  statement a):

{\it Let us assume that the norm of the Jordan product $a\circ b$ of two elements of  JB-algebra $\cal B$ is $\leq \epsilon.$ Then
\begin {equation}
\label{E}
||[Q_a,Q_b]||\leq  k( ||a||, ||b||) \sqrt\epsilon
\end {equation}
where $\epsilon \geq 0$ and $k$ is a polynomial function  }

To prove this statement we use the theorem that every JB-algebra can be considered  as a subalgebra of a direct sum of JC-algebra and purely exceptional JB algebra. (This follows from the fact that a JB-algebra $\cal B$ is a subalgebra of JBW-algebra ${\cal B}^{**}$ and every JBW-algebra is a direct sum of JW-algebra and purely exceptional JBW-algebra; see
\cite {WET} for  definitions and formulations and \cite {HO} for details and proofs). It is obvious that our statement is correct for a subalgebra if it is correct for the algebra. The statement is proven already for JC-algebra (\ref {a}). It remains to prove it for simple exceptional JB-algebra (then it is true for any purely exceptional algebra). To give the proof we modify slightly the considerations  given in  \cite {SHE} for $\epsilon =0.$ 

The only information we need  about simple exceptional JB-algebra (Albert algebra)  is the equation 
\begin {equation}
\label {CUB}
a^3=t(a)a^2-s(a)a+n(a)
\end {equation}
satisfied by any element $a$ of Albert algebra. (Here $t,s,n$ are linear, quadratic and cubic functions on Albert algebra.)

Notice first of all that for any JB-algebra $\cal B$
$$[Q_a,Q_b]= [R_{a^2}, R_{b^2}]+o_1$$
where $a,b\in \cal B$, $||o_1||\leq \epsilon f(||a||,||b||).$
(In \cite {SHE} this formula without the $o_1$ term is derived for $a\circ b=0.$ We use the same calculations, but instead of 
omitting terms containing $a\circ b$  we estimate them using $||a\circ b||\leq \epsilon.$ We apply the same procedure to the derivation of other formulas.) The next step is the formula
$$[Q_a,Q_b]c= [R_{a^2}, R_{b^2}]c+ o_1c=(a^2,c,b^2)+o_1c= -2(a^2\circ b,c,b) +o_1c$$
where $c\in \cal B$ and $(x,y,z)$ stands for the associator $(x\circ y)\circ z-x\circ (y\circ z).$

It follows from (\ref {CUB}) that
$$(a^2\circ b,c,b)=-s(a,b)(a,c,b)+o_2$$
where $s(a,b)$ denotes the bilinear form, corresponding to the quadratic form $s(a)$ and $||o_2||\leq \epsilon g(||a||,||b||)||c||. $ (See \cite {SHE} for the derivation in the case $\epsilon=0.$)  Substituting $c=a$ we obtain 
$$s(a,b)(a^2\circ b)=o_3$$
where $||o_3||\leq \epsilon h(||a||,||b||).$
We see that either $|s(a,b)|\leq ||o_3||^{\frac 1 2}$ or $||a^2\circ b||\leq ||o_3||^{\frac 1 2}.$
In the first case $$ ||[Q_a,Q_b]||\leq 2||o_3||^{\frac 1 2}||a|| ||b||+\epsilon (g(||a||,||b||)+||o_1|| .$$
In the second case $$ ||[Q_a,Q_b]||\leq 4 ||o_3||^{\frac 1 2} ||b||+||o_1||.$$

In both cases we obtain the estimate we need.

Let us consider again a commutative Lie group  (translation group) acting on JB-algebra $\cal B$ by means of automorphisms or, more generally, structural transformations $\alpha (\bx, \tau). $ We assume $(\alpha (\bx,\tau)\it a)\circ b$ is small for large $\bx$ . (If $\cal B$ is a JC-algebra corresponding to C*-algebra $\cal A$ this is equivalent to the asymptotic anticommutativity of elements $a,b$ in $\cal A.$) Then it follows from (\ref {E}) that {\it $Q_a$ and $Q_b$ asymptotically commute:}

{\it if $||(\alpha (\bx,\tau)\it a)\circ b||\leq \rho (\bx,\tau)$ then }$||[Q_{\alpha (\bx,\tau)\it a}, Q_b]||\leq k\sqrt {\rho(\bx,\tau)}.$

{\bf Acknowledgments} I am indebted to I. Shestakov for useful discussions.

\end {document}